\documentclass[conference]{IEEEtran}
\ifCLASSINFOpdf
\else
\fi
\usepackage{array}

\usepackage[shortcuts]{extdash}
\usepackage{mathtools}
\usepackage{mathpartir}
\usepackage{ amssymb } 
\usepackage{macros}
\usepackage{stmaryrd}
\usepackage{listings}
\usepackage{bigstrut}
\usepackage{color}
\usepackage{multirow}
\usepackage{courier}
\usepackage{paralist}
\usepackage[utf8]{inputenc}
\usepackage{todonotes}
\usepackage[boxed]{algorithm2e}
\SetCommentSty{textit}
\usepackage{arydshln}
\usepackage{pgfplots}
\usepackage{multirow}
\usepackage{ntheorem}
\usetikzlibrary{matrix,positioning}
\usepackage{float}
\usepackage[hyphens]{url}

\newcommand{\myparagraph}[1]{\smallskip \noindent{\bf #1}}
\newcommand{\mydef}{\triangleq}

{\theorembodyfont{\upshape}
\newtheorem{example}{Example}}
{\theorembodyfont{\upshape}
  \theoremheaderfont{\upshape}
  \theoremseparator{.}
 }

\usepackage{xspace}

\newcommand{\langname}{HTOL\xspace}
\newcommand{\langexpand}{Hyperlabel Test Objectives Language\xspace}

\newcommand{\card}[1]{|#1|}

\begin{document}

%
\title{Generic and Effective Specification of \\ Structural Test Objectives}

\author{\IEEEauthorblockN{Micha\"el Marcozzi, Micka\"el Delahaye,
S\'ebastien Bardin, Nikolai Kosmatov, Virgile Prevosto}
\IEEEauthorblockA{CEA, LIST, Software Reliability Laboratory\\         
91191 Gif-sur-Yvette, France\\
\textit{firstname.lastname@cea.fr}}
}


%


\maketitle

\begin{abstract}
{\noindent}
While a wide range of different and sometimes heterogeneous 
code-coverage criteria (a.k.a.~testing criteria, or adequacy criteria) have been proposed, 
there exists no generic formalism to describe them all, and
available test automation tools usually support only a small subset of them.
We introduce a  new specification language, called \langname
(\langexpand), 
providing a powerful generic mechanism to define a wide range of test objectives. 
\langname comes with a formal semantics,  and can encode all standard criteria but strong mutations.  
%
Besides specification, \langname\ is appealing  in the context of test automation as it allows to handle  criteria in a unified way.   
As a first practical application, 
we present a universal coverage measurement tool  supporting a wide range of standard criteria.      
Initial experiments demonstrate that the tool  
is practical and scales on realistic programs.  
\end{abstract}


%
\IEEEpeerreviewmaketitle

\section{Introduction}

\myparagraph{Context.} In current software engineering practice, testing \cite{myers,mathur,ZhuHM-97,ammann08} remains the primary approach to find bugs in a piece of code. We focus here on 
{\it white-box software testing}, in which the tester has access to the source code -- as it is the case for example in unit testing. 
 As testing all the possible program inputs is intractable in practice, the software testing community has notably defined \textit{code-coverage criteria} (a.k.a. \textit{adequacy criteria} or \textit{testing criteria}) \cite{ZhuHM-97,ammann08}, to select test inputs to be used. 
In regulated domains such as aeronautics, these coverage criteria are strict normative requirements that the tester must satisfy before delivering the software. 
In other domains, coverage criteria are recognized as a good practice for testing, and a key ingredient of test-driven development.

A coverage criterion fundamentally specifies a set of {\it test requirements} or {\it objectives}, which should be fulfilled by the selected test inputs.  
Typical requirements include for example covering all statements (statement coverage criterion) or all branches in the code (decision coverage criterion).  These requirements are essential to an automated white-box testing process, as they are used to guide  the selection of  new test cases, decide when testing should stop and assess the quality of a {\it test suite} (i.e., a set of test cases including test inputs).
In automated white-box testing, 
\textit{a coverage measurement tool} is used to establish which proportion of the requirements are actually covered by a given test suite, while a  \textit{test generation tool} tries to generate automatically a test suite satisfying the requirements of a given criterion.

\myparagraph{Problem.} Dozens of code-coverage criteria have been proposed in the literature \cite{ZhuHM-97,ammann08},  from basic control-flow or data-flow \cite{LK-83}  criteria 
to mutations \cite{DLS-78} and MCDC~\cite{CM-94},   offering notably different ratios between testing thoroughness and effort. 
However, from a technical standpoint, these criteria are seen as very dissimilar bases for automation, so that most testing tools (coverage measurement or test generation) 
are  restricted to a very small subset of criteria (cf.~Table~\ref{tab:covtools}) and that supporting a new  criterion is time-consuming. 
%
%
{\it Hence, the wide variety and deep sophistication of coverage criteria in academic literature is barely exploited in practice, and academic criteria have only a weak penetration into industry.} 


\myparagraph{Goal and challenges.} We intend to bridge the gap between the potentialities offered by the huge body of  academic work on (code-)coverage criteria on one side, 
 and their limited use in the industry on the other side. In particular, we aim at proposing  a {\it  well-defined and unifying specification mechanism for  these  criteria}, 
enabling a clear separation of concerns between  the precise declaration of  test requirements on one side, and the automation of white-box testing on the other side. 
This is a {\it fruitful} approach that has been successfully applied for example with SQL for databases and with temporal logics for model checking. This is also  a {\it challenging} task  as 
such a mechanism should be, at the same time: 
\begin{inparaenum}[(1)] 
\item
well-defined, 
\item expressive enough to encode test requirements from most existing criteria,
 and 
\item amenable to automation -- coverage measurement and/or test generation.  
\end{inparaenum}


\myparagraph{Proposal.} We introduce \textit{hyperlabels}, a generic specification language for white-box test requirements.  
%
%
%
%
Technically, hyperlabels are a major extension of {\it labels} proposed  by Bardin et {\it al.} \cite{bardin14}. While labels can express a large range 
of criteria \cite{bardin14} (including  a large part  \textbf{WM'} of weak mutations \cite{Howden-82},  and a weak variant of \textbf{MCDC} \cite{Pandita2010PexMCDC}),  they   are still too limited in terms of expressiveness, not being able for example to express 
strong variants of \textbf{MCDC} \cite{CM-94} or most dataflow criteria \cite{LK-83}.  In contrast, hyperlabels are able to encode {\it all  criteria from the literature} \cite{ammann08}   
 but  full mutations \cite{DLS-78,Howden-82}. 

%


Compared with similar previous attempts, hyperlabels try to find a sweetspot between genericity, specialization to coverage criteria and automation. Indeed, FQL~\cite{Holzer2010} cannot encode 
 \textbf{MCDC} or \textbf{WM'} but provides automatic test generation \cite{Holzer2008}, while temporal logics such as  HyperLTL or HyperCTL*~\cite{clarkson2014} 
are so expressive that automation faces significant scalability issues.  
Hyperlabels are both {\it necessary} and (almost) {\it sufficient} for expressing all interesting coverage criteria, and they seem to be amenable to {\it efficient} automation. 


\myparagraph{Contribution.} The four main contributions of this paper are:
\\
\textbf{1.}
We introduce a {\it novel taxonomy  of coverage criteria} (Section~\ref{cube}),  orthogonal  to both the standard classification \cite{ZhuHM-97} and the one 
by Ammann and Offutt \cite{ammann08}. Our classification is {\it semantical}, based on  the nature of the reachability constraints 
underlying 
a given criterion. 
%
This 
view  is sufficient for classifying  all existing criteria but mutations, and yields new insights into coverage criteria,  
emphasizing the complexity gap between  a given criterion and basic reachability.  
  A  visual representation of this taxonomy is proposed,  {\it the cube of coverage criteria}\footnote{By analogy to the $\lambda$-cube of functional programming.};    
\\
\textbf{2.}
We propose \langname (Hyperlabel Test Objective Language), a formal specification language for test objectives (Section \ref{hyperlabels}) based on {\it hyperlabels}. 
While labels reside in the cube origin, our language adds  new constructs 
for combining
(atomic) labels,  {\it allowing us to encode any criterion from the  cube taxonomy}.
We present \langname's syntax and give a formal semantics in terms of coverage.  
%
%
Finally, we give a few encodings of criteria beyond labels. Notably, \langname can express 
subtle differences between the variants of  \textbf{MCDC} (Section \ref{mcdc}); 
\\
\textbf{3.}
As a first application of hyperlabels, and in order to demonstrate their expressiveness, we provide in Section~\ref{encoding} a list of  encodings  for  
{\it almost all  code coverage criteria defined in the Ammann and Offutt  book~\cite{ammann08}}, including many criteria beyond labels (cf.~Table~\ref{fig:tabCritEnc}). The only missing criteria are strong mutations and full weak mutations, yet a large subset of weak mutations  can be encoded   \cite{bardin14}.  
\\
\textbf{4.}
As a second application of hyperlabels, and in order to demonstrate their practicality,  we present the design and implementation  of a universal and easily extensible 
code coverage measurement tool (Section \ref{tool}) 
  based on \langname. 
The tool already supports {\it in a unified way}  fourteen coverage criteria, including all criteria from Table \ref{tab:covtools} and six which are  beyond labels. 
%
We report on several experiments 
 demonstrating that the approach is 
efficient enough and scales well, both in terms of program  size and number of tests. 




\myparagraph{Potential impact and future work.} 
Hyperlabels provide a \textit{lingua franca} for defining, extending and comparing criteria in a clearly documented way, as well as a specification language for writing universal, extensible and interoperable testing tools. By making the whole variety and sophistication of academic coverage criteria much more easily accessible in practice, hyperlabels help bridging the gap between 
the rich body of 
academic results in criterion-based  testing and their limited use in the industry.
%
%
%
%
%



We intend to develop a test generation tool dedicated to hyperlabels in a middle term.  
Actually, automatic test generation can already be obtained by combining test generation for atomic labels \cite{bardin14}   with 
 coverage measurement for hyperlabels, yet a more dedicated technique is certainly  desirable.

\begin{table}[htbp]
 \caption{Criteria supported in a few  popular coverage tools}\label{tab:covtools} \vspace{-2mm}
  \def\X{$\checkmark$}%
  \scriptsize
\begin{center}
  \begin{tabular}{|l*{7}{|@{}>{\hspace*{3pt}}c<{\hspace*{3pt}}@{}}|}
    \hline
    
    Tool / Criterion & \tiny{FC} & \tiny{BBC} & \tiny{DC} & \tiny{CC} & \tiny{DCC} & \tiny{MCDC}  & \tiny{BPC}\bigstrut\\
    \hline
    \hline
    Gcov  & \X & \X & \X & & &  &\bigstrut\\
    \hline
    Bullseye  &\X & & & &  \X &   &\bigstrut\\
    \hline
    Parasoft & \X &  \X & \X & \X & & \X  & \X \bigstrut\\ 
    \hline
    Semantic Designs & \X &  & \X & & &  & \bigstrut \\
    \hline
    Testwell CTC++ & \X  & \X & & & \X & \X  & \bigstrut \\
    \hline
  \end{tabular}
\end{center}


FC: functions, BBC: basic blocks, DC: decisions, CC: conditions, DCC: decision condition, MCDC: modified decision condition,
BPC: basis paths  
\end{table}


\section{Overview}
\label{sec:overview}

We briefly sketch in Figure \ref{fig:overview} the workflow of our universal coverage measurement tool, in order to give an idea of how \langname\ helps 
to build test automation tools supporting a wide range of coverage criteria. 

{
\floatstyle{boxed}
\restylefloat{figure}

\begin{figure}[htbp]
\centering{\includegraphics[width=0.9\columnwidth]{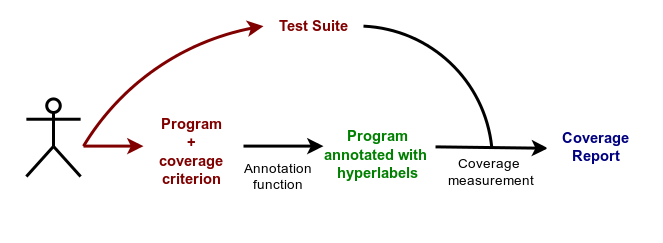}}%
\caption{Workflow of the universal coverage measurement tool} \label{fig:overview}
\end{figure}

}

The user provides a  program under test $P$ and a test suite $TS$, selects a coverage criterion $\mathbb{C}$  among a list of supported criteria and  obtains the coverage score of the test suite for the
  criterion $\mathbb{C}$.  
{\it Internally}, the program $P$ is first {\it automatically annotated} with hyperlabels {\it representing exactly the coverage objectives} defined by $\mathbb{C}$ (cf.~Figure~\ref{fig:EncodingExamples} for an example) -- we call 
{\it annotation function} (or labeling function) such a transformation, then coverage achieved by $TS$ is {\it measured  on the annotated program} 
and 
reported to the user.

From a developer's point of view, the analysis engine (here, coverage measurement) is written once and for all {\it (shared among criteria)}, and supporting a new criterion simply comes down to write a new annotation function 
{\it (shared among analysis engines)}.

\section{Background}

\subsection{Basics: Programs, Tests and Coverage} 



We give here a formal definition of coverage and coverage criteria, following \cite{bardin14}. 
Given a program $\Pg$ over a vector $V$ of $m$ input variables taking values in a domain
$\InputDom \triangleq \InputDom_1 \times \dots \times \InputDom_m$,
a \textit{test datum} $\TD$ for $\Pg$ is a valuation of $V$, i.e.\ $\TD \in \InputDom$. A \textit{test suite} $\TS \subseteq \InputDom$ 
is a finite set of test data. 
A (finite) execution of $\Pg$ over some $\TD$, denoted $\Pg(\TD)$, is a 
(finite) run $\sigma \triangleq \langle(\loc_0,\state_0),\dots,(\loc_n,\state_n)\rangle$ where the $\loc_i$ denote successive (control-)locations 
of $\Pg$ ($\approx$ statements of the programming language in which $\Pg$
is written) and the $\state_i$ denote the successive internal states of $\Pg$ ($\approx$ valuation of all global and local variables and of all memory-allocated structures)
after the execution of each $\loc_i$ ($\loc_0$ refers to the initial program state).

A test datum $\TD$ \textit{reaches} a location $\loc$ at step $k$ 
with internal state $\state$, denoted $\TD \covers_\Pg^{k} \langle \loc, \state\rangle$, if $\Pg(\TD)$ has the form
\mbox{$\sigma \cdot \langle \loc, \state\rangle \cdot \rho$}
where $\sigma$ is a partial run of length $k$.  When focusing on reachability, we omit $k$ and write $\TD \covers_\Pg \langle \loc, \state\rangle$. 

Given a test objective {\bf c}, we write  $\dt \covers_P \text{\bf c}$ if test datum $\dt$ covers {\bf c}.
We extend the notation for a test suite $\TS$ and a set of test objectives {\bf C}, writing   $\TS \covers_P \text{\bf C}$
when 
for any $\text{{\bf c}} \in \text{{\bf C}}$, there exists $\dt \in \TS$ such that $ \dt \covers_P \text{\bf c}$. 
%
{A \textit{(source-code based) coverage criterion} $\mathbb{C}$ is defined as a systematic way of deriving a set of test objectives $\text{{\bf C}}=\mathbb{C}(\Pg)$ for any program under test $\Pg$. 
A test suite $\TS$ satisfies (or achieves) a coverage criterion $\mathbb{C}$ if $\TS$ covers  $\mathbb{C}(\Pg)$.
When there is no ambiguity, 
we identify the coverage criterion $\mathbb{C}$ for a given program $\Pg$
with the derived set of test objectives $\text{{\bf C}}=\mathbb{C}(\Pg)$.}   

These definitions are generic and leave the exact definition of ``covering'' to the considered  coverage criterion.
%
%
For example,
test objectives derived from the Decision Coverage criterion are of the form $\text{\bf c} \mydef\ (\loc, \text{\tt cond})$ or $\text{\bf c} \mydef\  (\loc, \text{\tt !cond})$, where {\tt cond} is
 the condition of the branching statement at location $\loc$, and
  $\dt \covers_P \text{\bf c}$ if   $\dt$ reaches some  $(\loc, S)$ such that {\tt cond} evaluates to {\it true} (resp.~{\it false}) in $S$.

Finally, for a  test suite $\TS$ and a set {\bf C} of test objectives, the {\it coverage score} 
of $\TS$ w.r.t.  {\bf C}  is the ratio  
of the number of test objectives in $\text{{\bf C}}$ covered by $\TS$ to its cardinality $\card{ \text{{\bf C}}}$. 
The coverage score of $\TS$ w.r.t. a  coverage criterion $\mathbb{C}$ is then 
its coverage score w.r.t. the set $\text{{\bf C}}=\mathbb{C}(\Pg)$.

\subsection{A Quick Tour of  Coverage Criteria} \label{criteriaList}

A wide variety of criteria have been proposed in the literature  \cite{mathur,ammann08,ZhuHM-97}. We briefly review in this section the main  criteria used throughout the paper. 

\medskip


\textit{Control-flow graph coverage} criteria include basic block
coverage (\textbf{BBC}, equivalent to statement coverage), branch
coverage (\textbf{BC}) and several path-based criteria (where each one
specifies a particular set of paths to cover in the graph) such as
edge-pair (\textbf{EPC}), prime path (\textbf{PPC}), basis path
(\textbf{BPC}), simple/complete round trip
(\textbf{SRTC}/\textbf{CRTC}) and complete/specified path
(\textbf{CPC}/\textbf{SPC}) coverage.

\smallskip 

\textit{Call graph coverage} criteria include notably function coverage
(\textbf{FC}, all the call graph nodes, i.e.\ each program function
should be called at least once) and call coverage (\textbf{FCC}, all
the graph edges, i.e.\ each function should be called at least once
from each of its callers).

\smallskip

\textit{Data-flow coverage} \cite{LK-83} concerns checking that each value defined in the tested program is actually used, either by one of its possible uses (\textbf{all-defs}), 
or by all of them  (\textbf{all-uses}), or even along any of its def-use paths (\textbf{all-du-paths}).

\smallskip

\textit{Logic coverage} criteria focus on
exercising various truth value combinations for the
logical predicates (i.e.~branching conditions) of the tested program.  The most 
basic criteria here are  
decision coverage (both values for each predicate, \textbf{DC} --~equivalent to \textbf{BC}), 
(atomic) condition coverage (both values for each literal in each predicate, \textbf{CC}) 
and multiple condition coverage (all literal value combinations for each predicate, \textbf{MCC}). 
Advanced criteria include  \textbf{MCDC} \cite{CM-94} and its variants 
\cite{AmmannOH03,ammann08}  \textbf{GACC},  
\textbf{CACC} (masking MCDC) 
and \textbf{RACC} (unique-cause MCDC),  
as well as their inactive clause coverage counterparts  \textbf{GICC} 
and \textbf{RICC}.   
%
%
%
Other criteria consider 
the disjunctive normal form  of the predicates \cite[Chap.~3.6]{ammann08}, such as implicant coverage \textbf{IC}, 
unique true point coverage \textbf{UTPC} 
and corresponding unique true point and near
false point pair coverage \textbf{CUTPNFP} \cite{ChenL01}.  

\smallskip

Finally, in \textit{mutation coverage} \cite{DLS-78}, test requirements address the ability to detect that each of slight syntactic variants of the tested program (the {\em mutants}) behaves differently from the original code. In strong mutation coverage (\textbf{SM}), the divergence must be detected in the program outputs, whereas in weak mutation coverage (\textbf{WM}) \cite{Howden-82} the divergence must be detected just after the mutation.  
Both \textbf{SM} and \textbf{WM} are very powerful~\cite{ABL-05,OL-94}. Recently, Bardin {\it et al.} identified  side-effect-free weak mutations (\textbf{WM'}) \cite{bardin14,bardin15}  
as an expressive yet efficiently automatable fragment.

\subsection{Criterion Encoding with Labels}\label{labels}

In previous work, we have introduced \textit{labels}
\cite{bardin14}, a code annotation language to encode concrete test
objectives, and shown that several common coverage criteria
can be simulated by label coverage, i.e. given a program $P$ and a
criterion $\mathbf{C}$, the concrete test objectives instantiated
from $\mathbf{C}$ for $P$ can always be encoded using labels. As our
main contribution is a major extension of labels into hyperlabels, we
recall here basic results about labels. 

\myparagraph{Labels.} 
Given a program \Pg, a \textit{label} $\Lab \in \Labs$ is a pair $\langle \loc, \labpred \rangle$ where \loc is a location of P and \labpred is a predicate over the internal state at \loc, that is, such that:
\begin{inparaenum}[(1)]
  \item \labpred contains only variables and expressions
    (using in the same language as \Pg)
    defined at location \loc in \Pg, and
  \item \labpred contains no side-effect expressions.
\end{inparaenum}
There can be several labels defined at a single location, which can possibly
share the same predicate. More concretely, our labels can be compared to labels
in the C language, decorated with a pure C expression.

We say that  a test datum $\TD$ \textit{covers a label} $\Lab \mydef \langle\loc, \labpred \rangle$ in  $\Pg$, denoted $\TD \labcovers_\Pg \Lab$, if there is a state \state such that \TD 
reaches $\langle\loc,\state\rangle$ (i.e.\ $\TD \covers_\Pg \langle \loc, \state\rangle$) and \state satisfies \labpred. 
An \textit{annotated program} is a pair $\langle \Pg, \LabSet \rangle$ where $\Pg$ is a program and $\LabSet \subseteq \Labs$ is a set of labels for $P$.
Given an annotated program $\langle \Pg, \LabSet \rangle$, we say that a test suite \TS satisfies the \textit{label coverage criterion} (\textbf{LC}) for $\langle \Pg, \LabSet \rangle$, denoted $\TS \labcovers_{\langle \Pg, L \rangle} \mathbf{LC}$, if 
\TS covers every label of \LabSet (i.e. \ $\forall \Lab \in \LabSet : \exists \TD \in \TS: \TD \labcovers_\Pg \Lab$).

\myparagraph{Criterion Encoding.} 
Label coverage \textit{simulates a coverage criterion} $\mathbf{C}$
if any program $P$ can be {\em automatically} annotated with a set of
labels $L$ in such a way that any test suite $TS$ satisfies
$\mathbf{LC}$ for $\langle \Pg, \LabSet \rangle$ if and only if \TS covers
all the concrete test objectives instantiated from $\mathbf{C}$ for
$P$. 
We call  \textit{annotation} (or \textit{labeling})  \textit{function} such a procedure   
   automatically adding test objectives into a given program for a given coverage criterion.

\begin{figure}[htbp]
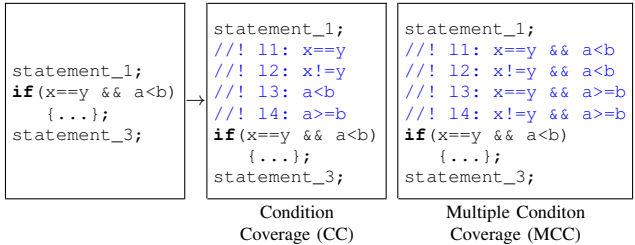

\centering%
\begingroup%
\noindent\scriptsize%
\setlength{\columnseprule}{0cm}%
\setlength{\tabcolsep}{0.7ex}%
\begin{tabular}{|c|@{}c@{}|c|@{}p{1pt}@{}|c|}
\cline{1-1}\cline{3-3}\cline{5-5}
&&&&\\[-0.6em]
\begin{lstlisting}
statement_1;
if(x==y && a<b)
   {...};
statement_3;
\end{lstlisting}
&$\rightarrow$&
\begin{lstlisting}
statement_1;
//! l1: x==y
//! l2: x!=y
//! l3: a<b
//! l4: a>=b
if(x==y && a<b)
   {...};
statement_3;
\end{lstlisting}
&&
\begin{lstlisting}
statement_1;
//! l1: x==y && a<b
//! l2: x!=y && a<b
//! l3: x==y && a>=b
//! l4: x!=y && a>=b
if(x==y && a<b)
   {...};
statement_3;
\end{lstlisting}
\\[-0.6em]
&&&&\\
\cline{1-1}\cline{3-3}\cline{5-5}
\multicolumn{1}{c}{}  &  \multicolumn{1}{c}{} & \multicolumn{1}{c}{\parbox[t]{8em}{\centering Condition \\ Coverage (CC)} \bigstrut}  & \multicolumn{1}{c}{}  & \multicolumn{1}{c}{\parbox[t]{9em}{\centering Multiple Conditon\\ Coverage (MCC)}} \\
\end{tabular}
\endgroup
\caption{Encoding of standard test requirements with labels (from \cite{bardin14})}
\label{fig:EncodingExamples}
\end{figure}

It is shown in \cite{bardin14}  that label coverage can notably simulate  basic-block coverage (\textbf{BBC}), branch coverage (\textbf{BC}) and decision coverage (\textbf{DC}), 
function coverage (\textbf{FC}), 
condition coverage (\textbf{CC}), decision condition coverage (\textbf{DCC}),  multiple condition coverage (\textbf{MCC})  as well as the  side-effect-free fragment of
 weak mutations (\textbf{WM'}). 
The encoding of \textbf{GACC} can also be deduced from  \cite{Pandita2010PexMCDC}. 
%
%
Figure \ref{fig:EncodingExamples} illustrates the simulation of some 
criteria with labels on sample code --  
that is,  the resulting annotated code automatically produced by the corresponding annotation functions.

The main benefit of labels is to {\it unify} the treatment of test requirements belonging to 
different classes of coverage criteria  in a transparent way, thanks to the {\it automatic insertion} of labels in the program under test. 

\myparagraph{Limits.} 
A label can only express the requirement that an assertion at a single location in the code must be covered by a single test execution. This is  not expressive enough to encode 
 the  test objectives coming from  
   path-based criteria, data-flow criteria, strong variants of \textbf{MCDC} or full mutations.  

\myparagraph{Our goal.} 
In this work, we aim at extending the expressive power of labels towards all the criteria presented in Section \ref{criteriaList} (except  \textbf{WM} and \textbf{SM}). The proposed extension should preserve the automation capabilities 
of labels.

\section{A New Taxonomy: The Cube} \label{cube}


We propose a new taxonomy for  code coverage criteria,  
based on the semantics of the associated  reachability problem\footnote{More precisely: the reachability problem of the test requirements associated to the coverage criterion.}. 
We take standard reachability constraints as a basis, and consider three orthogonal  extensions:
\begin{itemize}
\setcounter{enumi}{-1} 
  \item[{\bf Basis}] location-based reachability, constraining a single program location and a single test execution at a time, 
  \item[{\bf Ext1}] reachability constraints relating several executions of the same program (hyperproperties \cite{clarkson10}),  
  \item[{\bf Ext2}] reachability constraints along a whole 
    execution path (safety \cite{manna-safety}),  
  \item[{\bf Ext3}] reachability constraints involving choices between  several objectives. 
\end{itemize}

\begin{figure}[htbp]
  \quad 
  \centering{\includegraphics[width=\columnwidth]{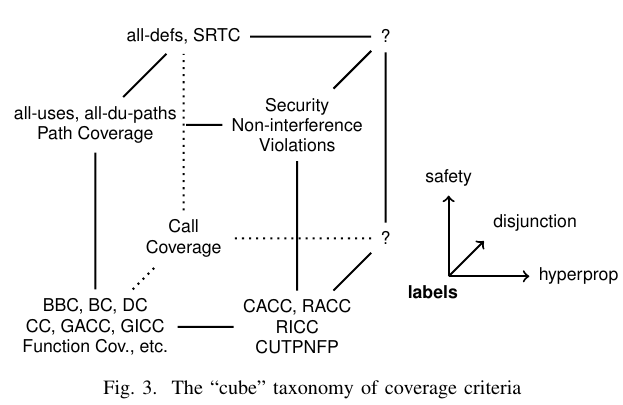}}%
\vspace{-2mm}
 \label{fig:cube}
\end{figure}

The basis  corresponds to  criteria that can be encoded with labels. Extensions 1, 2 and 3 can be seen as three euclidean axes that spawn from the basis 
 and add new capabilities to labels along three orthogonal directions. 
%
This gives birth to a visual representation of our taxonomy
as a cube, depicted in Figure \ref{fig:cube}, where each coverage
criterion from Section \ref{criteriaList} (but mutations) is arranged on one of the
cube vertices, depending on the  expressiveness of its associated reachability constraints. 
%
Intuitively, strong mutation falls outside the cube  because it relates two executions on {\it two programs}, the program under test and the mutant.  
Yet, we can  classify test objectives corresponding to the violation of  security properties such as non-interference (cf.~Example \ref{ex:non-int}, Section \ref{sec:first-examples}).   
%

This taxonomy is interesting in several respects.
First, it is {\it semantic}, in the sense that it refers to the reachability problems underlying the test requirements rather than to the artifact which the test requirements are drawn from. 
 In that sense  
it represents progress toward abstraction compared to the older taxonomies \cite{ammann08,ZhuHM-97}, the one of \cite{ammann08} being already more abstract than \cite{ZhuHM-97}. 
Second, it is very concise (only three basic parameters) and yet almost comprehensive, yielding new insights on criteria, through their distance to basic reachability. Interestingly, while many criteria require two extensions, we do not know of any criterion involving the three extensions. More generally, no
criterion seems to be using a disjunction of constraints over several
executions of the same program.



\section{Hyperlabels} \label{hyperlabels}


The previous section shows that our semantic taxonomy 
is suitable to 
represent
the whole set of coverage criteria we are interested in. 
Since labels correspond to basic reachability constraints, 
we seek to extend them in the three directions of axes 
in order to  build a universal test requirement description language. 
We detail here the principle, syntax and semantics of the proposed \langname language.

\subsection{Principles}\label{sec:principles}


\langname is based on labels \cite{bardin14} (referred to as {\it atomic} now)  to which we add five constructions, namely:   \textit{bindings}, \textit{sequences}, \textit{guards}, \textit{conjunctions} and \textit{disjunctions}. 
By combining these operators over atomic labels, one builds new objectives to be covered, which we call {\it hyperlabels}.

\begin{itemize}
\item  Bindings $\HypBind{\Lab}{\{v_1 \mapsfrom e_1; \ldots  \}}$ store in {\it meta-variable(s)} $v_1,\ldots$
the value of well-defined expression(s) $e_1,\ldots$
at the state at which atomic label $\Lab$ is covered;
 
 
\item Sequence 
$\Lab_1\HypSeqArrow{\phi}\Lab_2$ 
requires two atomic labels $\Lab_1$ and $\Lab_2$ to be covered 
sequentially by a single test run, constraining the whole path section
between them by $\phi$;

\item Conjunction $\Hyp_1 \HypConj \Hyp_2$ requires two hyperlabels $\Hyp_1, \Hyp_2$ to be covered by (possibly distinct) test cases, enabling to express {\em hyperproperties} 
about sets of tests;    

\item Disjunction $\Hyp_1 \HypDisj  \Hyp_2$ requires covering at least one of hyperlabels $\Hyp_1, \Hyp_2$. This enables  to simulate criteria involving disjunctions of objectives; 

\item Guard $\HypGuarded{\Hyp}{\hyppred}$ expresses a constraint ${\hyppred}$ over meta-variables
observed (at different locations and/or during distinct executions) when 
covering labels underlying $\Hyp$.

\end{itemize}

\subsection{First Examples}\label{sec:first-examples}

We present here a first few examples of criterion encodings using hyperlabels. They are presented in an informal way, a formal semantics of hyperlabels being given in 
Section \ref{sec:formal-def}. 

\begin{example}[MCDC]\label{ex:mcdc}
We start with  conjunction, bindings and guards. Consider the following code snippet:

\vspace*{-0.1cm}
{
\footnotesize
\centering\begin{lstlisting}
statement_0;
// loc_1
if (x==y && a<b) {...}; 
statement_2;
\end{lstlisting}
}

The (strong) \textbf{MCDC} criterion requires demonstrating that
each atomic condition $c_1 \mydef$ {\tt \small x==y} and $c_2
\mydef${\tt \small a<b} alone can influence the whole branch decision
$d \mydef c_1 \wedge c_2$.  
For $c_1$, it comes down to
providing two tests where the truth value of $c_2$ at $loc_1$ remains the same, 
while values of $c_1$ and $d$ change.
The requirement  for $c_2$ is symmetric.
This can be directly encoded with hyperlabels $h_1$ and $h_2$ as follows:
\vspace*{-0.25cm}
\begin{align*}
  l &\mydef
    \HypBind{(\loc_1, d)}{\{\mathit{c}_1 \mapsfrom \mathtt{\scriptstyle x==y}; \mathit{c}_2 \mapsfrom \mathtt{\scriptstyle a<b} 
   \}} \\
  l' &\mydef
   \HypBind{(\loc_1,\neg d)}{\{\mathit{c}'_1 \mapsfrom \mathtt{\scriptstyle x==y}; \mathit{c}'_2 \mapsfrom \mathtt{\scriptstyle a<b} 
  \}} \\
  h_1 &\mydef \HypGuarded{%
    l \HypConj l'
  }{%
    \mathit{c}_1 \neq \mathit{c}'_1 \wedge  \mathit{c}_2 = \mathit{c}'_2 
  }\\
  h_2 &\mydef \HypGuarded{%
    l \HypConj l'
  }{%
    \mathit{c}_1 = \mathit{c}'_1 \wedge  \mathit{c}_2 \neq \mathit{c}'_2  
  }\\
\end{align*}

\vspace*{-0.7cm}
$h_1$ requires that the test suite  reaches
{\small loc$_1$} twice (through the $\HypConj$ operator) -- with one or two tests  but different values for decision $d$.  
The values taken by  the atomic conditions  
   when {\small loc$_1$} is reached 
 are bound  (through \HypBind{}{}\!\! ) to metavariables  $c_1, c_2$ (first 
execution) and   $c'_1, c'_2$ (second one). Moreover, 
  these recorded values must satisfy the guard  $\mathit{c}_1 \neq \mathit{c}'_1 \wedge  \mathit{c}_2 = \mathit{c}'_2$, 
meaning that $c_1$ alone can influence the decision.
Similarly, $h_2$ ensures the desired test objective for $c_2$.
 \end{example}

\begin{example}[Call coverage]\label{ex:call-cov}
Let us continue by showing the interest of the disjunction operator. Consider the following code snippet where 
{\tt \small f} and {\tt \small g} are two functions.  

{
\footnotesize
\centering\begin{lstlisting}
int f() {
if (...) { /* loc_1 */ g(); }
if (...) { /* loc_2 */ g(); }}
\end{lstlisting}
}

The function call coverage criterion (\textbf{FCC}) requires a test case going from {\tt \small f} to {\tt \small g}, i.e.~passing either through $loc_1$ or $loc_2$. 
This is exactly represented by hyperlabel $h_3$ below: 
\[h_3 \mydef (loc_1, \text{\it true}) \HypDisj (loc_2, \text{\it true})\]
\end{example}

\begin{example}[all-uses]\label{ex:all-uses}
  We illustrate now the sequence operator
  $\HypSeqArrow{\cdot}$. Consider the following code snippet.

{
\footnotesize
\centering\begin{lstlisting}
/* loc_1 */ a := x; 
if (...) /* loc_2 */ res := x+1;
else /* loc_3 */ res := x-1;
\end{lstlisting}
}

In order to meet the \textbf{all-uses} dataflow criterion  for the definition of variable {\tt \small a} at line $loc_1$, a test suite 
 must  cover the two def-use paths from $loc_1$ to  $loc_2$ and to $loc_3$. 
These two objectives are represented by 
hyperlabels $h_4 \mydef (loc_1, \text{\it true}) \HypSeqArrow{} (loc_2, \text{\it true}) $ and
$h_5 \mydef (loc_1, \text{\it true}) \HypSeqArrow{} (loc_3, \text{\it true}) $. 
\end{example}

\begin{example}[Non-interference]\label{ex:non-int}
  Last, we present a more demanding example that involves bindings,
  sequences and guards.  {\it Non-interference} is a strict security
  policy model which prescribes that information does not flow between
  sensitive data (\textit{high}) towards non-sensitive data
  (\textit{low}).  This is a typical example of hypersafety
    property \cite{clarkson10,clarkson2014}. 
Hyperlabels  can
express the violation of such a property in a straighforward manner.  Consider 
the code snippet below.  

{\footnotesize
\begin{lstlisting}
int flowcontrol(int  high,  int  low) {
  // loc 1
  {...}
  // loc 2
  return res; }
\end{lstlisting}
}

Non-interference is violated here if and only if two executions with the same {\tt \small low} input exhibit different output ({\tt \small res}) -- because it would mean that 
a difference in the {\tt \small high} input is observable. This can be encoded with hyperlabel $h_6$:    
\begin{align*}
  l_1 &\mydef
    \HypBind{(\loc_1,\text{\it true})}{\{\mathit{lo} \mapsfrom \mathtt{\scriptstyle low}\}} \rightarrow
    \HypBind{(\loc_2,\text{\it true})}{\{\mathit{r} \mapsfrom \mathtt{\scriptstyle res}\}}\\
  l_2 &\mydef
    \HypBind{(\loc_1,\text{\it true})}{\{\mathit{lo}' \mapsfrom \mathtt{\scriptstyle low}\}} \rightarrow
    \HypBind{(\loc_2,\text{\it true})}{\{\mathit{r}' \mapsfrom \mathtt{\scriptstyle res}\}}\\
  h_6 &\mydef \HypGuarded{l_1 \HypConj l_2}{lo = lo' \wedge r \neq r'}\\
\end{align*}
\end{example}

\subsection{Formal definition} \label{sec:formal-def}


\myparagraph{Syntax.} The syntax is given in Figure~\ref{fig:hyp-syntax}, 
where:
\begin{itemize}
  \item $\Lab \mydef \langle \loc, \labpred \rangle \in \Labs$ is an atomic label.
  \item $\bindings \in \Bindings_\loc$ is a partial mapping between arbitrary metavariable names $v \in \HypNames$ and well-defined expressions $e$ at the program location \loc ;
  \item $l, l_1, \cdots , l_i , \cdots ,l_n$ are atomic labels with bindings; 
  \item $\phi_i$ is a predicate over the metavariable names defined in the bindings of labels $l_1,\dots,l_i$, over the current program location $pc$ ($\approx$ program counter) and over the variable names defined in all program locations that can be executed in a path going from $\loc_i$ to $\loc_{i+1}$.
  \item $\Hyp, \Hyp_1, \Hyp_2 \in \Hyps$ are hyperlabels;
  \item $\hyppred$ is a predicate over the set $\HypNM(\Hyp)$ 
of \textit{\Hyp-visible names} 
(i.e. metavariable names \textit{guaranteed} to be recorded by \Hyp's bindings),
defined as follows:
  \begin{align*}
  \HypNM(\HypBind{\Lab
	 }{\bindings}) & \mydef
     \text{all the names defined in }\bindings \\
   \HypNM([ l_1 \xrightarrow{\phi_1} \cdots~ l_n ]) & \mydef 
     \HypNM(l_1) \cup \cdots \cup \HypNM(l_n)\\             
   \HypNM(\HypGuarded{\Hyp}{\hyppred}) &  \mydef  \HypNM(\Hyp)\\
   \HypNM(\Hyp_1 \HypConj \Hyp_2) &  \mydef  \HypNM(\Hyp_1) \cup \HypNM(\Hyp_2)\\
   \HypNM(\Hyp_1 \HypDisj \Hyp_2) &  \mydef  \HypNM(\Hyp_1) \cap \HypNM(\Hyp_2);
  \end{align*}
\end{itemize}

{

\floatstyle{boxed}
\restylefloat{figure}

\begin{figure}[htbp]

  \small\centering\[\begin{aligned}
    \Hyp ~ \texttt{\textbf{::=}} \qquad & l && \text{label} \\
			 \texttt{\textbf{|}} \quad & [ l_1 \xrightarrow{\phi_1} \texttt{\textbf{\{}} l_i \xrightarrow{\phi_i} \texttt{\textbf{\}}}^\texttt{\textbf{*}}~l_n ] && \text{sequence of labels}\\             
              \texttt{\textbf{|}}  \quad & \HypGuarded{\Hyp}{\hyppred} && \text{guarded hyperlabel}\\
              \texttt{\textbf{|}}  \quad & \Hyp_1 \HypConj \Hyp_2 && \text{conjunction of hyperlabels}\\
              \texttt{\textbf{|}}  \quad & \Hyp_1 \HypDisj \Hyp_2 && \text{disjunction of hyperlabels}\\
              & && \\
                  l ~ \texttt{\textbf{::=}} \qquad & \HypBind{\Lab}{\bindings} && \text{atomic label with bindings}\\
              & && \\
                 \bindings ~  \texttt{\textbf{::=}} \qquad & \{v_1 \mapsfrom e_1; \ldots  \}  &&  \text{bindings}   \\ 
  \end{aligned}\]


\caption{Syntax of Hyperlabels} \label{fig:hyp-syntax}
\end{figure}

}

\myparagraph{Well-formed hyperlabels.} In general, a name can be bound multiple times in a single hyperlabel, which
would result in ambiguities when evaluating guards. To prevent this issue, we
define in Figure~\ref{fig:wf} a well-formed predicate $\HypWF{\cdot}$ over hyperlabels. In the remaining part of this paper,  we will
only consider well-formed hyperlabels.

{

\floatstyle{boxed}
\restylefloat{figure}

\begin{figure}[htbp]
  \centering
  \begin{mathpar}
    \inferrule{\forall i,j, i\neq j \Rightarrow v_i\neq v_j }
    {\HypWF{\HypBind{\Lab}{\{v_1 \leftarrow e_1;...; v_n\leftarrow e_n \}}}}

    \inferrule{\HypWF{\Hyp}}{\HypWF{\HypGuarded{\Hyp}{\hyppred}}}
    \\
    \inferrule
    {\forall i,j,\, i\neq j\Rightarrow \HypNM(l_i)\cap \HypNM(l_j)=\emptyset}
    {\HypWF{[ l_1 \xrightarrow{\phi_1} \cdots~ l_n ]}}
    \\
    \inferrule{\HypWF{h_1}\\\HypWF{h_2}\\\HypNM(l_1)\cap \HypNM(l_2)=\emptyset}
    {\HypWF{h_1\HypConj h_2}}
    \\
    \inferrule{\HypWF{h_1}\\\HypWF{h_2}\\\HypNM(l_1)=\HypNM(l_2)}
    {\HypWF{h_1\HypDisj h_2}}
  \end{mathpar}
  \caption{Well-formed hyperlabels}
  \label{fig:wf}
\end{figure}

}

In particular, on well-formed hyperlabels,  
$\HypNM$ is compatible with distributivity of $\HypConj$ and $\HypDisj$. For
instance, if we have $\HypWF{h}$ with $h \mydef h_1 \HypConj (h_2 \HypDisj h_3)$, then,
with $h' \mydef (h_1 \HypConj h_2)\HypDisj(h_1\HypConj h_3)$, we have $\HypWF{h'}$ and
$\HypNM(h)=\HypNM(h')$.



\myparagraph{Semantics.} \langname is given a semantics in terms of  {\it coverage} and {\it execution traces}, as was done for atomic labels~\cite{bardin14}.  
This kind of semantic is not tied to syntactic elements of the program under test, allowing for example to express  {\bf WM'}.

A primary requirement for covering hyperlabels is to capture execution states
into the variables defined in bindings.
For that, we introduce the notion of \textit{environment}. An environment $\HypEnv\in \HypEnvs$ is a partial mapping between names and values, that is,
$\HypEnvs \mydef \HypNames\nrightarrow\HypValues$.
Given an execution state \state at the program location \loc and some bindings $\bindings \in \Bindings_\loc$, the \textit{evaluation} of $\bindings$ at state $\state$, noted $\llbracket \bindings \rrbracket_\state$ is an environment $\HypEnv \in \HypEnvs$ such that $\HypEnv(v) = val$ iff $\bindings(v)$  evaluates to $val$ considering the execution state $\state$.

We can now define \textit{hyperlabel coverage}. A test suite \TS \textit{covers} a hyperlabel $\Hyp \in \Hyps$, noted $\TS \hypcovers_\Pg \Hyp$, if there exists some environment $\HypEnv \in \HypEnvs$ such that the pair $\langle \TS, \HypEnv \rangle $ covers \Hyp, noted $\langle \TS, \HypEnv \rangle \hypcovers_\Pg \Hyp$, defined by the inference rules of Figure \ref{fig:HypSemRules}.  An \textit{annotated program} is a pair $\langle \Pg, H \rangle$ where $\Pg$ is a program and $H \subseteq \Hyps$ is a set of hyperlabels for $P$. 
Given an annotated program $\langle \Pg, H \rangle$, we say that a test suite \TS satisfies the \textit{hyperlabel coverage criterion} (\textbf{HLC}) for $\langle \Pg, H \rangle$, noted $\TS \hyplabcovers_{\langle P, H \rangle} \mathbf{HLC}$ if the test suite \TS covers every 
hyperlabel from $H$ (i.e.\ $\forall h \in H
 : \TS \hypcovers_\Pg \Hyp$).
\begin{figure*}[htbp]
  \centering\framebox{\parbox{0.98\textwidth}{%
\begin{mathpar}
  \inferrule[Label]{
    \TD \in \TS \\
    \TD \covers_\Pg^{k} \langle\loc,\state\rangle\\
    s \vDash \labpred \\
    \HypEnv \supseteq \llbracket B \rrbracket_s
  }{
  \TD \covers_{\HypEnv}^{k}  \HypBind{\langle \loc, \labpred \rangle 
   }{\bindings}
 \\
      \langle \TS,\HypEnv \rangle \hypcovers_\Pg 
   \HypBind{\langle \loc, \labpred \rangle  
   }{\bindings}
    }
\and
\inferrule[Guard]{
    \langle \TS,\HypEnv \rangle \hypcovers_\Pg \Hyp \\
    \HypEnv \vDash \hyppred 
  }{
    \langle \TS,\HypEnv \rangle \hypcovers_\Pg \HypGuarded{ \Hyp }{ \hyppred } 
  }
  \and
\inferrule[Conjunction]{
    \langle \TS,\HypEnv \rangle \hypcovers_\Pg \Hyp_1 \\
    \langle \TS,\HypEnv \rangle \hypcovers_\Pg \Hyp_2 
  }{
    \langle \TS,\HypEnv \rangle \hypcovers_\Pg \Hyp_1 \HypConj \Hyp_2 
  }\\
  \inferrule[Disjunction Left]{
    \langle \TS,\HypEnv \rangle \hypcovers_\Pg \Hyp_1 
  }{
    \langle \TS,\HypEnv \rangle \hypcovers_\Pg \Hyp_1 \HypDisj \Hyp_2 
  }
  \and
  \inferrule[Disjunction Right]{
    \langle \TS,\HypEnv \rangle \hypcovers_\Pg \Hyp_2 
  }{
    \langle \TS,\HypEnv \rangle \hypcovers_\Pg \Hyp_1 \HypDisj \Hyp_2 
  }
  \and
  \inferrule[Sequence]{
    \TD \in \TS \\
    \forall i \in \left[1,n\right], ~ \TD \covers_\HypEnv^{k_i} l_i \\
    \forall i \in \left[1,n-1\right], ~ k_i < k_{i+1} \\\\
    \forall i \in \left[1,n-1\right], ~ \forall j \in \left]k_i,k_{i+1}\right[, ~
       (loc_j,s_j)=P(t)_{j} \, \wedge \, \phi_i(\HypEnv,loc_j,s_j)
  }{
    \langle \TS,\HypEnv \rangle \hypcovers_\Pg [ l_1 \xrightarrow{\phi_1} \texttt{\textbf{\{}} l_i \xrightarrow{\phi_i} \texttt{\textbf{\}}}^\texttt{\textbf{*}}~l_n ] 
  }
\end{mathpar}%
\begin{scriptsize}%
Naming convention: %
$\TS$ test suite;
\HypEnv hyperlabel environment;
$\Hyp, \Hyp_1, \Hyp_2$ hyperlabels;
$\hyppred$ hyperlabel guard predicate;
$n$ positive integer;
$l_1,\dots,l_n$ atomic labels with bindings;
$\TD$ test datum;
$k, k_1,\dots, k_n$ execution step numbers;
$\loc_j, \loc$ program locations;
$\state_j, \state$ execution states;
$P(t)_{j}$ the $j$-th step of the program run $P(t)$ of $P$ on $t$;
$\phi_1,\dots,\phi_n$ predicates over sequences of labels;
\labpred label predicate;
\bindings hyperlabel bindings.
\end{scriptsize}
}}
  \caption{Inference rules for hyperlabel semantics}\label{fig:HypSemRules}
\end{figure*}

The criterion simulation introduced for labels \cite{bardin14} is generalized to hyperlabels. Hyperlabel coverage simulates a coverage criterion $\mathbf{C}$ if any program $P$ can be automatically annotated with a set of hyperlabels $H$, so that, for any test suite $TS$, $TS$ satisfies $\mathbf{HLC}$ for $\langle \Pg, H \rangle$ iff \TS fulfills all the concrete test objectives instantiated from $\mathbf{C}$ for $P$.

\myparagraph{Disjunctive Normal Form.}  \label{sec:dnf}
Any well-formed hyperlabel $h$ can be rewritten into a \textit{disjunctive normal form} (DNF), 
i.e a  {\it coverage-equivalent} hyperlabel  $h_{dnf}$ arranged as  a disjunction 
$h_{dnf} \mydef c_1 + \cdots + c_i + \cdots + c_n$ 
of  {\it guarded conjunctions}  $c_i \mydef \langle ls_1^i \HypConj \dots \HypConj ls_p^i ~|~ \psi(B_{ls_1^i},\cdots,B_{ls_p^i})\rangle$ 
over  atomic labels or sequences. 
%
%
%
%
%
%
%
%
%
The   equivalence between $h$ and $h_{dnf}$ is stated as
   $$\forall~ TS \subseteq D ~\forall~ \HypEnv \in \HypEnvs, \langle \TS, \HypEnv \rangle \hypcovers_\Pg \Hyp \Leftrightarrow \langle \TS, \HypEnv \rangle \hypcovers_\Pg \Hyp_{hnf}.$$


 DNF normalization is an important step of our coverage measurement algorithm (cf.~Section \ref{sec:cover-algo}). 


\subsection{Advanced Examples}




\subsubsection{Playing with MCDC variants}\label{mcdc}  
Example \ref{ex:mcdc} provides an encoding of the strongest version of \textbf{MCDC} (a.k.a.~\textbf{RACC}). Yet, weaker variants exist. 
Encoding them into hyperlabels helps clarify the subtle 
differences between those variants.

\medskip 

\textbf{GACC} (General Active Clause Coverage) is the weakest variant of  \textbf{MCDC}. It is also the sole variant encodable with atomic labels \cite{Pandita2010PexMCDC}.  
Let us assume that we have a predicate $p$   composed of $n$ atomic conditions $c_1,\dots, c_n$. 
\textbf{GACC} requires that for each  $c_i$, the test suite triggers two distinct executions of the predicate: one   where $c_i$ is true, one   where $c_i$ is false, and both such that the truth value of $c_i$ impacts the truth value of the whole predicate. Yet, it is not required that switching the value of $c_i$ is indeed feasible, and the two 
executions do not have to be correlated.  
Going back to the code snippet of Example~\ref{ex:mcdc}, 
\textbf{GACC} requirement for $c_1$ can be simulated by 
$l_3$ and $l_4$, where $d(x,y)$ denotes decision $d$ (cf.~Example~\ref{ex:mcdc})   where $c_1$ and $c_2$ are replaced by $x$ and $y$.  

$l_3 \mydef (loc_1, c_1 \wedge d(true,c_2) \neq d(false,c_2) )$ 

$l_4 \mydef (loc_1, \neg c_1 \wedge d(true,c_2) \neq d(false,c_2) )$


\medskip \smallskip 

\textbf{CACC} (Correlated Active Clause Coverage), or
masking \textbf{MCDC} is stronger than \textbf{GACC}. It
includes every requirement from \textbf{GACC} and additionally requires that for each clause $c_i$, the two executions  are such that if $p$ is true (resp.\ false) in the first one, 
then it is false (resp. true) in the second one.  
\textbf{CACC} cannot be encoded into atomic labels because of this last requirement that correlates the two executions together. 
Yet, it can be encoded with hyperlabels.
Using the same code as in Example~\ref{ex:mcdc}, \textbf{CACC} requirement for 
$c_1$ can be simulated by the following hyperlabel $h_7$, built on the two atomic labels $l_3$ and $l_4$ 
defined for \textbf{GACC}: 
 
$h_7 \mydef  \HypGuarded{ \HypBind{l_3}{\{r \HypBindArrow{}{}d\}}  \HypConj   \HypBind{l_4}{\{r' \HypBindArrow{}{}d\}}     }{r \neq r'}   $

\medskip

\subsubsection{More DataFlow criteria} \label{ad}

The \textbf{all-defs} coverage criterion requires that each definition of a variable must be connected to {\it one of its} uses. The criterion adds a disjunction of objectives to the 
\textbf{all-uses} criterion. 
Going back to Example~\ref{ex:all-uses}, the \textbf{all-defs} requirement for the definition of variable {\tt \small a} at line $loc_1$ can 
be simply simulated by hyperlabel $ h_8 \mydef h_4 \HypDisj h_5$.


\medskip 

\label{cells} 
Finally, data-flow criteria can be refined to consider
the {\bf definition and use of single array cells}, while the standard
approach considers arrays as a whole. Indeed,
the index of the accessed cells may
not be known statically, making it impossible to relate defs and uses, as
well as to define def-free paths without dynamic information. For
example, in the following code, the path from $\loc_1$ to $\loc_3$ is
a valid du-path iff $i = k \neq j$, which cannot be known statically:

{\footnotesize 
\begin{lstlisting}
int foo(int i,int j,int k){
  /* loc_1 */ a[i] = x;
  /* loc_2 */ a[j] = y;
  /* loc_3 */ z = a[k] + 1; }
\end{lstlisting}
}

With hyperlabels, we just have to add bindings to the atomic labels for saving the values of $i$ and $j$ and use the guard operator to force them being equal. Encoding for the previous example is given below,
with $pc$ the current line of code:

$l_5 \mydef (\loc_1,true) $ \qquad  $l_6 \mydef (\loc_3,true)  $   

\smallskip 

$h_9 \mydef \HypGuarded{\HypBind{l_5}{\{v_1 \mapsfrom i\}} \xrightarrow{\substack{pc=loc_2\\\Rightarrow j \neq v_1}} \HypBind{l_6}{\{v_2 \mapsfrom k\}}}{v_1=v_2}$

\medskip

\subsubsection{Path-based Criteria}\label{sec:path-based-criteria}

Most test objectives coming from path-based criteria have a straightforward
encoding with the $\HypSeqArrow{}$ operator, typically  \textbf{complete path coverage} (for a finite number of paths). 
%
%
A few criteria also require  operator $\HypDisj$  for choices between  paths, e.g.~\textbf{simple round trip coverage}. 
%

\section{Extensive  Criteria Encoding} \label{encoding}


As a first application of hyperlabels, we perform an extensive literature review and we try to  encode all coverage criteria with hyperlabels. Especially, we have been able to encode all  criteria from the Ammann and Offutt book \cite{ammann08}, but strong mutations and full weak mutations.
Indeed, these two criteria really require the ability to run tests on variants
of the original program, whereas \langname does not modify the code itself.
These results are summarized in Table \ref{fig:tabCritEnc}, where we also specify which criteria can be expressed by atomic labels alone, and  the required hyperlabel operators otherwise.

\newlength{\MyNegSpaceA}
\setlength{\MyNegSpaceA}{-0.2cm}
\newcommand{\MyNegHSpaceA}{\hspace{\MyNegSpaceA}}

\begin{table}[htbp]
  \caption{Simulation of criteria 
from 
\cite{ammann08}. 
}\label{fig:tabCritEnc} \smallskip 
\begin{scriptsize}
  \centering
  \begin{tabular}{|l|p{0.3cm}@{}|p{0.28cm}@{}|p{0.28cm}@{}|p{0.28cm}@{}|p{0.28cm}@{}|p{1.2cm}@{}|}
  \cline{2-7}
  \multicolumn{1}{c|}{} & \multicolumn{5}{c|}{\centering\textbf{\bigstrut Encodable by}} & {\textbf{\bigstrut See Sec.}} \\ 
  \multicolumn{1}{c|}{} &  \multirow{3}{*}{\rotatebox{90}{\makebox[6mm]{\textbf{labels}}}} &  \multicolumn{4}{c|}{\centering \textbf{hyperlabels}}  & {\textbf{\bigstrut or ref.}}\\
  \multicolumn{1}{c|}{} &  &  \multicolumn{4}{c|}{\centering \textbf{using}}  & \\
  \multicolumn{1}{c|}{} & & \centering  \MyNegHSpaceA $\xrightarrow{\phi}$ & \centering \MyNegHSpaceA $| \psi \rangle$ & \centering \MyNegHSpaceA $\cdot$ & \centering \MyNegHSpaceA $+$ & \\ 
  \hline 
  \textbf{Control-flow graph coverage} &  &  &  &  &  &  \\ 
Statement,\,Basic-Block,\,Branch & $\checkmark$ &  &  &  &  &  \cite{bardin14} \\ 
\textit{Path coverage:} & & & & & & \\
\quad EPC,\,PPC,\,CRTC,\,CPC,\,SPC  & & $\bullet$ &  &  &  & \ref{sec:path-based-criteria} \\ 
\quad Simple Round Trip coverage & & $\bullet$ &  &  & $\bullet$ & \ref{sec:path-based-criteria}  \\ 
\textbf{Call-graph coverage} &  &  &  &  &  &  \\
Function coverage (all nodes) & $\checkmark$ &  &  &  &  & \ref{labels} \\ 
Call coverage (all edges) & &  &  &  & $\bullet$ & \ref{ex:call-cov} \\ 
\textbf{Data-flow coverage} &  &  &  &  &  &  \\ 
All Definitions (all-defs) &  & $\bullet$ &  &  & $\bullet$ &  \ref{ad}  \\ 
~~ + array cell definitions & & $\bullet$ & $\bullet$  &  & $\bullet$ &  \ref{cells}  \\ 
All Uses (all-uses) &  & $\bullet$ &  &  &  & \ref{ex:all-uses} \\ 
~~ + array cell definitions & & $\bullet$ & $\bullet$  &  &  &  \ref{cells}  \\ 
All Def-Use Paths (all-du-paths) &  & $\bullet$ &  &  &  & \ref{sec:path-based-criteria} \\ 
~~ + array cell definitions &  & $\bullet$ & $\bullet$  &  &  &  \ref{cells}  \\ 
  \textbf{Logic expression coverage} &  &  &  &  &  &  \\
BBC, CC, DCC, MCC                      & $\checkmark$ &  &  &  &  &  \cite{bardin14} \\ 
\textit{MCDC variants}:  & & & & & & \\
\quad GACC, GICC  & $\checkmark$ &  &  &  &  &  
\ref{mcdc},\,\cite{Pandita2010PexMCDC} \\ 
\quad CACC, RACC, RICC & &  & $\bullet$ & $\bullet$ &  & \ref{mcdc} \\
\textit{DNF-based criteria}: & & & & & & \\
\quad IC, UTPC & $\checkmark$ &  &  &  &  &   \\ 
\quad CUTPNFPPC & &  & $\bullet$ & $\bullet$ &  &  \\ 
  \textbf{Mutation coverage} &  &  &  &  &  &  \\
  Side-effect-free Weak Mut. & $\checkmark$ &  &  &  &  &  \cite{bardin14} \\
    \cline{2-6}
  (Full) Weak Mut., Strong Mut. & \multicolumn{5}{c|}{\centering not encodable} &  \\
      \cline{2-6}
\hline 
  \end{tabular} 

\smallskip 

$\checkmark$: expressible by atomic labels  \qquad $\bullet$: required hyperlabel operators

\vspace{-3mm}
\end{scriptsize}
\end{table}

Interestingly, many criteria fall beyond the scope of atomic labels, and many also require combining two or three \langname operators. This is a strong {\it a posteriori} evidence that 
the language of hyperlabel is both {\it necessary} and (almost) {\it sufficient} to encode state-of-the-art coverage criteria. 
{\it Detailed encodings are available on the companion website}\footnote{Companion website: \url{http://icst17.marcozzi.net}}.

\section{Universal Coverage Measurement Tool} \label{tool}

As a second application, we describe  a {\it universal} coverage measurement tool, built on \langname, following the view of Section \ref{sec:overview} and the   
philosophy of LTest~\cite{Bardin2014TAP}. 
The two basic building blocks are: 
(1) a coverage measurement procedure for test suites on 
programs annotated with hyperlabels, and 
(2) pre-defined (hyperlabeling) annotation functions for 
  standard criteria (cf.~Section~\ref{encoding}). 

This prototype is the first coverage measurement tool able to handle {\it all} coverage criteria from \cite{ammann08} (but strongest mutation variants) in a {\it unified way}. 
Fourteen criteria are supported so far -- their annotation functions are provided: six based on  hyperlabels (\textbf{CACC}, \textbf{RACC}, \textbf{FCC}, \textbf{BPC}, \textbf{all-defs} and \textbf{all-use}),  
plus eight based on atomic labels\footnote{Namely: \textbf{FC}, \textbf{BBC}, \textbf{DC}, \textbf{CC}, \textbf{DCC}, \textbf{MCC}, \textbf{GACC} and \textbf{WM'}.}.   
Supporting  a new coverage criterion amounts to implement its  annotation function.  

While our coverage measurement algorithm runs in worst-case exponential time considering the whole HTOL expressiveness,  experiments 
demonstrate that the tool is efficient enough on existing coverage criteria, and scales well with both program size and number of tests.  
%


\subsection{Computing the coverage of a test suite} \label{sec:cover-algo}

Given an annotated program $\langle P, H \rangle$ and a test suite $TS$, our coverage measurement algorithm follows  three steps.
\begin{LaTeXdescription}
\item[normalization] First, each hyperlabel $h \in H$ is rewritten into its {\it disjunctive normal form} (cf.~Section \ref{sec:formal-def}). 
\item[harvesting] Second, each test case $t$ from $TS$ is run on $P$. Every atomic label and label sequence covered during the run is saved on-the-fly, together with the environment (values of metavariables) that instantiates the label's bindings at the coverage points.  
\item[consolidation] Third, the collected coverage information is  propagated within the syntax tree (in DNF) of every $h \in H$, in order to establish if $TS$ covers $h$ or not.  
\end{LaTeXdescription} 

These steps are now described in more details. 

\myparagraph{Normalization.}\label{dnfHL} 
As stated in Section \ref{sec:dnf}, any  (well-formed) hyperlabel $h$ can be rewritten into an equivalent hyperlabel $h_{dnf}$ in  disjunctive normal form. 
This form of labels is both very convenient for coverage measurement and
very common in practice. This is done by applying the
rewrite rules of Figure~\ref{fig:dnf-rewriting-rules} bottom-up from the
leaves of the hyperlabel tree.
The proof of equivalence between $h$ and $h_{dnf}$ can easily be obtained by 
induction on $h$.





{

\floatstyle{boxed}
\restylefloat{figure}

\begin{figure}[htbp] 
\scriptsize
\begin{mathpar}

\inferrule{\ }{l \dnfrew \langle l | \text{\it true}  \rangle } \and 
\inferrule{\ }{s \dnfrew \langle s | \text{\it true}  \rangle } \and
\inferrule{h \dnfrew \sum_i \langle \pi_i | {\psi}_i \rangle  }{\langle h | {\psi} \rangle \dnfrew    \sum_i \langle \pi_i | {\psi}_i  \land {\psi} \rangle } 
\and
\inferrule{h^L \dnfrew \sum_i \langle \pi_i^L | {\psi}_i^L \rangle  \qquad h^R \dnfrew \sum_j \langle \pi_j^R | {\psi}_j^R \rangle    }{ h^L + h^R \dnfrew    \sum_i \langle \pi_i^L | {\psi}_i^L \rangle + \sum_j \langle \pi_j^R | {\psi}_j^R \rangle  }
\and
\inferrule{h^L \dnfrew \sum_i \langle \pi_i^L | {\psi}_i^L \rangle  \qquad h^R \dnfrew \sum_j \langle \pi_j^R | {\psi}_j^R \rangle    }{ h^L \cdot h^R \dnfrew    \sum_i \sum_j  \langle \pi_i^L \cdot \pi_j^R | {\psi}_i^L \land {\psi}_j^R \rangle   }\and

\text{\it notation:\quad} \pi \mydef  l \cdot \ldots \cdot s \cdot  
\end{mathpar}
\caption{Rewriting hyperlabel into DNF}
\label{fig:dnf-rewriting-rules}\label{algoDNF}
\end{figure}

}

\myparagraph{Environment harvesting.} Once hyperlabels in DNF have been obtained, 
  each test $t$ from the suite $TS$ is run on $P$, and the  coverage information for basic labels, sequences and binding values is collected. 
Note that we need to store all possible binding values encountered along the execution of $t$, not just the first one. 
While this is easy for atomic labels, sequences must be treated with care, as there are some non deterministic choices there.  
Due to space limitations, we do not describe this point here,  common  in runtime monitoring 
{\it A detailed description is available on the companion website}\footnote{Companion website: \url{http://icst17.marcozzi.net}}. 


\myparagraph{Consolidating coverage result.} Once the coverage information for basic labels, sequences and binding values is fully collected, we can compute the whole 
hyperlabel-coverage information. This is straightforward on DNF hyperlabels: 
\begin{itemize}
\item atomic labels and sequences with no  guard are covered iff they have been covered in the harvesting step;  

\item a guarded conjunction  $c \mydef \langle ls_1 \HypConj ... \HypConj  ls_p ~|~ \psi(B_{ls_1},...B_{ls_p})\rangle$ is covered iff 
each label or sequence $ls_{j}, j \in 1..p$ is saved as covered in $E$ and there is at least one set of  environments $\HypEnv_j \in E$ (one for every $ls_j$ with bindings) 
 such that $\psi(\HypEnv_1, \cdots, \HypEnv_p)$ is true;  

\item a disjunction $h_{hnf} \mydef c_1 + \cdots + c_i + \cdots + c_n$ is covered iff at least one of the $c_i$ is covered.

\end{itemize}

In practice, the tool tries every possible combination of $\HypEnv_j$ from $E$ for every $c_i$, until it finds one which makes $\psi$ true (in which case $TS$ covers $h$) or proves that none exists (in which case $h$ is not covered by $TS$). Note that under the assumption that $P$ terminates for all test cases in 
$TS$, this algorithm is both correct and complete with respect to \langname's
semantics, that is, it will find a combination covering
a well-formed hyperlabel $\Hyp$ iff
$\TS \hypcovers_P \Hyp$. This
can be shown by induction on the number of disjunctions and conjunctions
in the normalized form of $h$.

\myparagraph{Optimizations.} We first preprocess  hyperlabels under consideration  in order to remove all unused metavariables appearing in bindings.   Then, during harvesting, we ensure that each binding is recorded only once, avoiding duplicated values. Finally, we perform conjunction and disjunction evaluation in a lazy way, in order to avoid unnecessary combinatorial reasoning 
on  guarded conjunctions.

\myparagraph{About complexity.} 
The algorithm presented so far runs in 
worst-case exponential time, mainly because of three factors:
\begin{inparaenum}[(1)]
\item normalization may yield an exponential-size hyperlabel,
\item consolidation for guarded conjunctions may lead to checking a number
of solutions exponential in the size of the conjunction,  
and 
\item monitoring sequences of labels may
include harvesting a number of environments exponential in the length of the
considered run.
\end{inparaenum}

Yet, in practice, our algorithm appears to {\it perform well on existing classes of testing requirements}  (cf.~Section \ref{sec:xp}). 
Here are a few explanations. 
First, criteria as encoded in the previous sections are naturally in DNF. 
Second, the critical parameters indicated above are  strongly limitated in existing criteria: 
conjunctions   of length 2; sequences  of length 2 or without  bindings;  
small  domains of metavariables (boolean). 
In that setting, complexity becomes polynomial.

\subsection{Implementation} \label{sec:implem}

We have implemented a basic hyperlabel support in LTest~\cite{Bardin2014TAP}, 
an open-source all-in-one testing platform for C programs,  
developed as a  Frama-C \cite{FRAMAC} plugin. 
LTest is built on standard labels, and provides (labeling) annotation functions to automatically encode requirements from common coverage criteria, coverage measurement, automatic coverage-oriented test generation~\cite{bardin14} 
and automatic detection of infeasible requirements~\cite{bardin15}. LTest relies on PathCrawler~\cite{WilliamsMM04ASE} for test generation and on Frama-C for static analysis.   

Our prototype extends LTest in two aspects. First, we provide an 
{\it hyperlabeling mechanism}, together with (hyperlabeling) annotation functions dedicated to the supported criteria.   
Second, we have implemented coverage measurement for hyperlabels. 
%

\subsection{Experimentations} \label{sec:xp}

\myparagraph{Objective.}
We want to assess the practical applicability of our universal coverage measurement tool, at least for unit testing.  

\noindent \quad [\textbf{\textit{RQ 1}}] Is the proposed unified approach practical and efficient enough? More precisely, how does the tool scale with large test suites on criteria beyond labels? 


\smallskip 

\myparagraph{Protocol.}
We consider 13 C functions split up into 3 groups:
\begin{itemize}
\item 5 small witness functions, mainly from Siemens \cite{sir}, 
Verisec \cite{verisec} \& MediaBench \cite{mediabench}, as already used in \cite{bardin14};  
\item 5 functions from OpenSSL 1.0.2 \cite{openssl}, a $250kloc$ open-source application. We focus on modules of about $1kloc$. 
\item 3 functions from SQLite 3.13 \cite{sqlite}, a $215kloc$ open-source application. We focus on  modules of a few $kloc$. 
\end{itemize}
The C files automatically annotated with HTOL test objectives are available on companion website 
\url{http://icst17.marcozzi.net}.

\noindent \quad For [\textbf{\textit{RQ 1}}],  a set of up to 10,000 test cases is randomly generated for each C function. Our tool is successively run with an increasing number of these unit test cases, which can also be downloaded {\it from the companion website}.
Each tool run is repeated 7 times. First, tests are executed without measurement (baseline), and then  measuring coverage  for the \textbf{CC} and \textbf{GACC} label-encodable criteria (witness). 
Second, tests are measured for the \textbf{CACC}, \textbf{RACC}, \textbf{FCC} and \textbf{all-defs} criteria, which involve the five operators from hyperlabels. 
%
%
%
%
%
All  experiments are performed under Ubuntu Linux 14.04 on an Intel Core i7-4712HQ CPU at 2.30GHz, 
with 16GB of RAM.


\smallskip 

\begin{figure}[htbp]
\centering{\includegraphics[width=0.9\columnwidth]{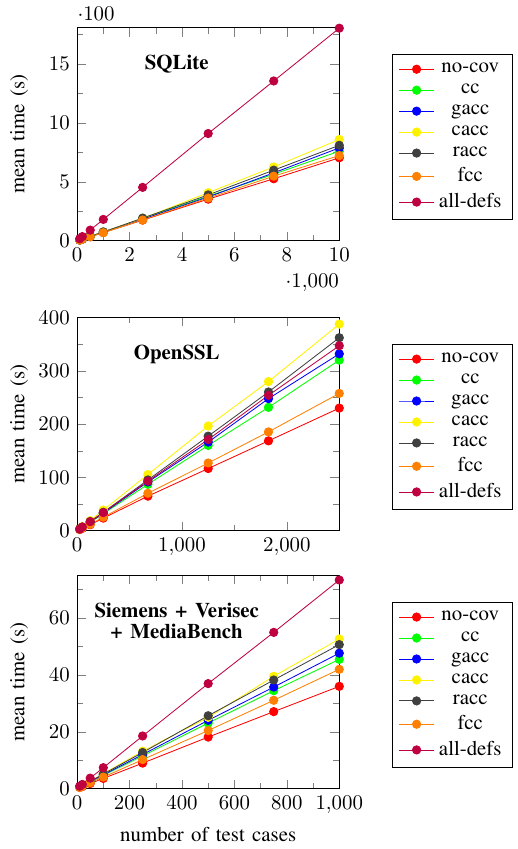}}%
\caption{Scalability of Coverage Measurement}
\label{scal-bench}
\end{figure}

\myparagraph{Results and discussion.} Our main results are presented in Figure \ref{scal-bench}. 
Detailed results are part of the annexes and {\it full results are available on the companion website}.  \vspace{0.1cm}

\noindent \quad [\textbf{\textit{RQ 1}}] 
Figure \ref{scal-bench} plots, for each criterion and the baseline (no-cov), the mean measurement time for all programs, as a function of the test suite size. 
We can notice that: (1) the measurement time grows linearly with the number of test cases, (2) the time overhead is very reasonable for all criteria but \textbf{all-defs} (between 1.1x and 2x), 
and still not so high for \textbf{all-defs}  (between 2x and 4x), and (3) these results hold on the three benchmarks, regardless of program size.  
%
%
Note that \textbf{all-defs} yields a tangible time overhead on some programs, due to the higher number of test objectives that our implementation defines. 
However, many of these objectives are trivial or redundant, which could be detected using some control-flow  analysis in an optimized version of the tool. 

{\it Conclusion.} These results indicate that upgrading labels with hyperlabels makes it possible to build an (almost) universal coverage measurement tool, without losing practical applicability. The measurement time for criteria beyond labels is acceptable and remains linear with the size of the test suite. Moreover, as our tool implementation is not optimized, there is still room for a strong reduction of coverage measurement time, when using the approach in a more industrial context. \vspace{0.1cm}

\section{Related Work} \label{sec:related}

The two closest works to ours are {\it labels} \cite{bardin14} and {\it
  FQL}. Since the difference with labels has already been presented 
(Sections~\ref{labels} and~\ref{sec:principles}, Table~\ref{fig:tabCritEnc}), we focus here 
on FQL.

\myparagraph{Specification of white-box coverage criteria.}  
 The \textit{Fs\-hell Query Language} (FQL) by Holzer {\it et al.} \cite{Holzer2010} for test suite specification and the associated Fshell \cite{Holzer2008} tool represent  the closest work to ours. FQL enables encoding code coverage criteria into an extended form of regular expressions, whose alphabet is composed of elements from the control-flow graph of the tested program. Fshell  takes advantage of an off-the-shelf model-checker to automatically generate from a C program  a test suite satisfying a given FQL specification. 

The scope of criteria that can be encoded in FQL is incomparable with the one offered by \langname, as FQL  handles complex safety-based test requirements but no hyperproperty-based requirement.  
Moreover, FQL is limited to syntactic elements of the program under analysis.  
As a consequence, FQL  cannot encode neither  \textbf{MCDC} nor \textbf{WM'}.  

Yet, FQL  offers the interesting ability to encode, in an elegant and standardized way,  generic coverage criteria (independently of any concrete program),   
where \langname encodes concrete test objectives (i.e.\ particular instantiations of coverage criteria for  given  programs).  
%
%
Note also that  FShell  provides automatic test generation, while we focuses on coverage measurement for now. 

\myparagraph{Specification of model-based coverage criteria.} Blom {\it et al.} \cite{Blom04} proposes to specify test objectives 
on extended finite state machines (EFSMs) 
as observer automata with parameters, while Hong {\it et al.} \cite{Hong2002} considers CTL temporal logic.  
Formal encodings have also been   proposed for several model-based coverage criteria in different other formalisms, like set theory \cite{set93}, graph theory \cite{graph90}, predicate logic \cite{predicate96,predicate00}, OCL \cite{friske2008composition} and Z \cite{Vilkomir2008}.  However, for each formalism, the scope of supported criteria is   limited to safety-based criteria, 
with no support of hyperproperties.

\myparagraph{Coverage objectives and hyperproperties.} 
%
Hyperproperties~\cite{clarkson10} are  properties over several  traces of a system.  
Testing  hyperproperties is a rising issue, notably in the frame of security  \cite{kinder15}. However, research in the topic still remains exploratory. 
Rayadurgam {\it et al.} \cite{Rayadurgam03} suggests that \textbf{MCDC} can be encoded with temporal logics, by writing the formulas for a self-composition of the tested model with itself. The paper reports that model-checking the obtained formulas rapidly faces scalability issues. 
Clarkson {\it et al.}  \cite{clarkson2014} introduces HyperLTL and HyperCTL*, which are extensions of temporal logics for hyperproperties, as well as an associated model-checking algorithm. This work makes no reference to test criterion encoding, but the proposed logics could be used to provide \cite{Hong2002} with the ability to encode criteria like \textbf{MCDC}. However, the complexity results and first experiments  \cite{clarkson2014} indicate that the approach  faces strong scalability limits. 
\langname being {\it a priori} less generic (yet, sufficient in practice), it is likely to be  more amenable to {\it efficient} automation.  
In future work, we intend to explore how \langname formally compares to HyperLTL and HyperCTL*. 
%
%



\myparagraph{Test description languages.}  Some languages have been designed to support the implementation of test harnesses at the program (TSTL \cite{Groce:2015}, UDITA \cite{gligoric10}) or model (TTCN-3 \cite{Grabowski03}, UML Testing Profile \cite{Schieferdecker03}) level. A test harness is the helper code that executes the testing process in practice, which notably includes test definition, documentation, execution and logging. These languages offer general primitives to write and execute easily test suites, but independently of any explicit reference to a coverage criterion.

\myparagraph{Coverage measurement tools.} Code coverage is used extensively in
the industry.  As a result, there exists a lot of testing tools that embed some
sort of coverage measurement.  For instance, in 2007, a
survey~\cite{yang2009survey} found ten tools for programs written in the C
language:
Bullseye~\cite{bullseye},
CodeTEST, 
Dynamic~\cite{dynamic},
eXVantage, 
Gcov (part of GCC)~\cite{gcov},
Intel Code Coverage Tool~\cite{intelcodecov},
Parasoft~\cite{parasoft},
Rational PurifyPlus,
Semantic Designs~\cite{semanticdesigns},
TCAT~\cite{tcat}. 
To this date, there are even more tools, 
such as COVTOOL, LDRAcover~\cite{ldracover},
and Testwell CTC++~\cite{testwellctcpp}.  

As a rule of thumb, these tools support a limited number
of test criteria in a hard-coded, non-generic manner.   Table~\ref{tab:covtools} (Section \ref{tab:covtools})  
summarizes implemented criteria for some popular tools. 
Our prototype already supports all these criteria  in a generic and extensible way, plus seven other criteria (cf.~Section \ref{sec:implem}).    
%
%
%
However, to be fair, code coverage tools also aim at causing as little
overhead as possible.  In contrast,  
as a first step, we only aim at getting a reasonable overhead. 

\section{Conclusions} \label{conclusion}

To sum up, \langname proposes a unified framework for describing and
comparing most existing test coverage criteria. 
This enables in particular implementing generic tools that can be used for a
wide range of criteria. We propose as a first application a universal coverage measurement tool, with an overhead  sufficiently low to
not be a concern in practice. 
%
Future work includes the efficient lifting of automatic 
test generation technologies to \langname. 






\bibliographystyle{IEEEtran}
\bibliography{IEEEabrv,main}

\end{document}